# Evolution of two bulk-superconducting phases in Sr$_{0.5}$RE$_{0.5}$FBiS$_2$ (RE: La, Ce, Pr, Nd, Sm) by external hydrostatic pressure effect

Aichi Yamashita[1], Rajveer Jha[1], Yosuke Goto[1], Akira Miura[2], Chikako Moriyoshi[3], Yoshihiro Kuroiwa[3], Chizuru Kawashima[4], Kouhei Ishida[4], Hiroki Takahashi[4] and Yoshikazu Mizuguchi[1,*]

[1]Department of Physics, Tokyo Metropolitan University, 1-1 Minami-Osawa, Hachioji, Tokyo, 192-0397, Japan

[2]Faculty of Engineering, Hokkaido University, Kita-13, Nishi-8, Kita-ku, Sapporo, Hokkaido 060-8628, Japan

[3]Graduate School of Advanced Science and Engineering, Hiroshima University, 1-3-1 Kagamiyama, Higashihiroshima, Hiroshima 739-8526, Japan

[4]Department of Physics, College of Humanities and Sciences, Nihon University, Setagaya, Tokyo 156-8550, Japan

Polycrystalline samples of Sr$_{1-x}$RE$_x$FBiS$_2$ (RE: La, Ce, Pr, Nd, and Sm) were synthesized via the solid-state reaction and characterized using synchrotron X-ray diffraction. Although all the Sr$_{0.5}$RE$_{0.5}$FBiS$_2$ samples exhibited superconductivity at transition temperatures ($T_c$) within the range of 2.1–2.7 K under ambient pressure, the estimated superconducting volume fraction was small. This indicated the non-bulk nature of superconductivity in these samples under ambient pressure. A dramatic evolution of the bulk superconducting phases was achieved on applying an external hydrostatic pressure. Near pressures below 1 GPa, bulk superconductivity was induced with a $T_c$ of 2.5–2.8 K, which is termed as the low-$P$ phase. Moreover, the high-$P$ phase ($T_c$ = 10.0–10.8 K) featuring bulk characteristics was observed at higher pressures. Pressure-$T_c$ phase diagrams indicated that the critical pressure for the emergence of the high-$P$ phase tends to increase with decreasing ionic radius of the doped RE ions. According to the high-pressure X-ray diffraction measurements of Sr$_{0.5}$La$_{0.5}$FBiS$_2$, a structural phase transition from tetragonal to monoclinic also occurred at approximately 1.1 GPa. Thus, this phase transition indicates a pressure-induced superconducting–superconducting transition similar to the transition in LaO$_{0.5}$F$_{0.5}$BiS$_2$. Bulk superconducting phases in Sr$_{0.5}$RE$_{0.5}$FBiS$_2$ induced by the external hydrostatic pressure effect are expected to be useful for evaluating the mechanisms of superconductivity in BiCh$_2$-based superconductors.

## Introduction

The discovery of superconductivity in BiS$_2$-based compounds such as Bi$_4$O$_4$S$_3$ and $RE$O$_{1-x}$F$_x$BiS$_2$ has triggered numerous studies focusing on identifying new superconductors featuring a higher transition temperatures ($T_c$) and elucidating the mechanisms of superconductivity in BiS$_2$-based compounds. Due to the similarity between the crystal structures of cuprates and iron-based compounds[1,2], BiS$_2$-based compounds have been considered as a type of layered superconductors. Thus far, various types of BiS$_2$-based compounds have been synthesized by replacing different blocking layers[3-14], such as the Bi$_4$O$_4$(SO$_4$)$_{1-x}$ or ($RE$O) ($RE$: La, Ce, Pr, Nd, and Sm) layers. By manipulating the layered structures, the replacement of ($RE$O) blocking layers with the SrF blocking layers enabled the development of a new class of BiS$_2$-based Sr$_{1-x}RE_x$FBiS$_2$ ($RE$: La, Ce, Pr, Nd, and Sm) compounds[12, 15–21]. The parent compound is SrFBiS$_2$, which is a semiconductor with a band gap. The substitution of Sr$^{2+}$ with $RE^{3+}$ induces electron carriers in the BiS$_2$ layer, and filamentary superconductivity appears at approximately 2.8 K in La-, Pr-, and Ce-doped compounds[12, 17–19].

Notably, some BiS$_2$-based compounds do not exhibit bulk superconductivity even after electron doping. To induce bulk superconductivity in BiS$_2$-based compounds, various chemical substitutions have been attempted[22–24]. Among these, the iso-valent substitutions, such as Nd$^{3+}$ substitutions for La$^{3+}$ or Se$^{2-}$ substitutions for S$^{2-}$, were found to be effective for inducing bulk superconductivity. Based on systematic structural analyses, we have revealed that the in-plane chemical pressure is one of the essential parameters that facilitate the emergence of bulk superconductivity in Bi$Ch_2$-based ($Ch$: S, Se) systems[25,26].

Regarding the superconducting mechanisms, unconventional pairing mechanisms were suggested based on the recent first principal calculations[27–29]. In addition, angle-resolved photoemission spectroscopy (ARPES) has revealed a strongly anisotropic superconducting gap or the possibility of sign changes in the superconducting gap[30]. Recently, we have reported the absence of isotope effects on $T_c$ in LaO$_{0.6}$F$_{0.4}$BiSSe[31], and the two-fold-symmetric in-plane anisotropy of magnetoresistance in the superconducting states in tetragonal LaO$_{0.5}$F$_{0.5}$BiSSe with a four-fold-symmetric in-plane structure[32]. This is likely related to nematic superconductivity states, which have been observed in several unconventional superconductors[33]. These findings indicate the emergence of unconventional superconducting states in the $RE$(O,F)Bi$Ch_2$ system. Therefore, in order to further investigate the superconducting mechanisms of Bi$Ch_2$-based compounds, it is essential to develop a new Bi$Ch_2$-based system.

In this study, we focus on the Sr$_{0.5}RE_{0.5}$FBiS$_2$ ($RE$: La, Ce, Pr, Nd, and Sm) system. Based on electrical resistivity measurements conducted under high pressures, a considerable increase in $T_c$ was observed in previous pressure experiments[15–17]. The highest $T_c$ in Sr$_{0.5}RE_{0.5}$FBiS$_2$ is approximately 10 K, and the pressure dependences of $T_c$ exhibits a sharp increase at the critical pressure (~ 1 GPa). It was suggested that the significant increase in $T_c$ under pressure in BiS$_2$-based systems was associated with a structural transition from tetragonal to monoclinic, which was revealed via structural analyses of LaO$_{0.5}$F$_{0.5}$BiS$_2$ and EuFBiS$_2$[34,35]. Although there has been no further report on the pressure-induced superconductivity in $Sr_{0.5}RE_{0.5}FBiS_2$, we assumed that a similar structural transition occurs in its case as well. Considering these facts and assumptions regarding the $Sr_{0.5}RE_{0.5}FBiS_2$ system, we performed magnetization measurements on the system under high pressures in order to obtain information regarding the superconducting characteristics of both the low-$P$ and high-$P$ phases. Here, we present the evolution of two bulk superconducting phases: the low-$P$ and high-$P$ phases. Furthermore, we characterized the crystal structure of $Sr_{0.5}La_{0.5}FBiS_2$ under high pressures; accordingly, a possible structural transition from tetragonal to monoclinic was observed at approximately 1.1 GPa.

## Results

**Sample characterization and physical properties at ambient pressure**

Fig. 1 depicts the powder synchrotron X-ray diffraction (XRD) patterns for $Sr_{0.5}La_{0.5}FBiS_2$. For $Sr_{1-x}RE_xFBiS_2$ ($RE$ = Ce, Pr, Nd, and Sm), see supplemental Fig. S1 (a-d). The crystal structure of the obtained samples was well refined using the Rietveld method. They were well refined using a tetragonal structure with the $P4/nmm$ space group. Small impurity peaks due to $REF_3$ ($RE$: La, Ce) and $Bi_2S_3$ were also detected in the case of the Pr, Nd, Sm-based samples. As shown in Fig. 2, we observed that the lattice constant $a$ decreased with decreasing $RE$ ionic radius; however, the lattice constant $c$ increased under these conditions. The obtained values were in agreement with previous reports[12, 16]. The chemical composition ratios of the samples were determined using energy dispersive X-ray spectroscopy (EDX). These results showed that the chemical compositions of the obtained samples were in reasonable agreement with the nominal compositions (Table 1). The electrical resistivity of the samples at ambient pressure was measured to be 1.6 K (Fig. 3). The semiconducting behaviour in all the samples was observed; moreover, the $\rho$-$T$ curve indicated a slight increase in $\rho$ on cooling, which implied that the conduction electrons were weakly localized due to the in-plane local disorder in the $BiS_2$ layer. A superconducting transition was observed at $T_c$ = 2.7, 2.7, 2.6, 2.6, and 2.1 K for $RE$ = La, Ce, Pr, Nd, and Sm, respectively (Fig. 3 inset). In addition, a superconducting transition was also observed in the temperature dependence of magnetization, as plotted in Fig. 4. The $T_c$ estimated based on magnetization was in agreement with that obtained based on the resistivity measurements. This is the first study to report on the observation of superconductivity in $Sr_{0.5}Nd_{0.5}FBiS_2$ and $Sr_{0.5}Sm_{0.5}FBiS_2$ under ambient pressure. However, in these samples, the estimated shielding volume fraction was less than 6%. To obtain bulk superconductivity, which is essentially important for describing intrinsic superconducting properties, we applied external pressure on the samples.

**External pressure effect**

Fig. 5 (a) shows the temperature dependences of the magnetization of $Sr_{0.5}La_{0.5}FBiS_2$ when increasing the applied pressure to 1.15 GPa. The superconducting transition temperature of approximately 2.7 K (low-$P$ phase) remained almost unchanged up to 0.84 GPa; alternatively, there was an evident increase in the shielding volume fraction. This result indicates that the external pressure effectively enhances the shielding volume fraction, which corresponds to the bulk nature of the superconducting states in the $Sr_{0.5}La_{0.5}FBiS_2$ samples. A remarkable increase in $T_c$ up to $T_c^{max}$ = 10.8 K (high-$P$ phase) was observed at $P$ > 0.95 GPa. This maximum value of $T_c$ is comparable to that obtained for a high-$P$ phase of $LaO_{0.5}F_{0.5}BiS_2$[34,37–39] and $Sr_{0.5}La_{0.5}FBiS_2$[15,40] through resistivity measurements under high pressure. The enhancement in the shielding volume fraction was also observed in the high-$P$ phase, which was achieved by increasing the applied pressure to the maximum pressure, without a noticeable change in the $T_c$. Moreover, a similar trend was observed in the low-$P$ region. The drastic increase in $T_c$ is likely related to the structural transition from a tetragonal to a monoclinic phase, which was suggested for the related superconductor $LaO_{0.5}F_{0.5}BiS_2$ based on the XRD under high pressure[34]. The pressure dependence of $T_c$ is summarized in Fig. 5 (f); the light blue, blue, and pink regions indicate the filamentary superconductivity, bulk superconductivity in the low-$P$ phase, and bulk superconductivity in the high-$P$ phase, respectively. The low-$P$ phase shifts to the high-$P$ phase when a pressure slightly exceeding the critical pressure ($P_c$) for the high-$P$ phase is applied. In regard to this, a phase diagram was created using the shielding volume fraction of 20% or more as a bulk state in order to discuss the phase transition. In the pressure range of 0–0.84 GPa, $T_c(P)$ remains almost constant (~ 2.7 K). The bulk superconductivity of the sample was induced by the pressure for the low-$P$ phase region; eventually, the high-$P$ phase region emerged at approximately $P$ = 0.95 GPa with the bulk superconducting states. The $T_c(P)$ for the high-$P$ phase region is almost constant (10.8 K). The temperature dependence of magnetization under pressure and the pressure dependence of $T_c$ phases for all the samples are summarized in Fig. 5 (a-e) and (f-j), respectively. The high-$P$ phase of $Sr_{0.5}Sm_{0.5}FBiS_2$ was not observed up to 1.28 GPa, which is nearly the upper limit of the pressure measurement apparatus used in this experiment. This is the first report offering evidence of the bulk nature of two different (low- and high-$P$) phases of the $Sr_{1-x}RE_xFBiS_2$ ($RE$: La, Ce, Pr, Nd, and Sm) superconductors when subjected to pressure.

## Discussion

In this section, we discuss the relationship between external pressure effects, chemical pressure effects, and evolution of superconductivity in $Sr_{0.5}RE_{0.5}FBiS_2$. On comparing the evolutions of superconductivity for $RE$ = Ce–Sm and that for $RE$ = La, a slight decrease in $T_c$ for the low-$P$ phase was observed with increasing pressure for the $RE$ = Ce, Pr, Nd and, Sm samples. This deviation in the behaviour might be explained by the effect of coexistence of both the external as well as chemical pressures[25]. Moreover, we observed that the $T_c$ of the high-$P$ phase varied slightly with the $RE$ ionic radius. On the contrary, the $P_c$ for the high-$P$ phase increased with decreasing $RE$ ionic radius. Specifically, assuming a coordination number of 8, $P_c$ was roughly estimated as 0.95, 1.11, 1.17, and 1.33 GPa for $La^{3+}$ (with an ionic radius of 1.16 Å), $Ce^{3+}$ (with an ionic radius of 1.14 Å), $Pr^{3+}$ (with an ionic radius of 1.13 Å), and $Nd^{3+}$ (with an ionic radius of 1.11 Å), respectively. The tendency of the increase in $P_c$ and the decrease in $T_c$ with a decrease in $RE$ ionic radius was also observed for $REO_{0.5}F_{0.5}BiS_2$ compounds[38,39,40]; specifically, the $T_c$ varied from approximately 10 K to 6 K, and a significantly higher $P_c$ was required with decreasing $RE$ ionic radius. These observations highlight the relationship between the structural transition to monoclinic and the packing density in the $BiS_2$ conducting layers[34]. It is possible that the larger space along the BiS conduction plane with a larger $RE$ ionic radius possibly transmits the pressure, thereby inducing structural transition. As shown in Fig. 6, $T_c^{mag}$ of all samples for the low- and high-$P$ phases were plotted with respect to the lattice constant $a$. $T_c$ of the high-$P$ phase was defined as the highest $T_c$ for the range of applied pressures, whereas $T_c$ of the low-$P$ phase was defined as the value of $T_c$ when the applied pressure is lower than the pressure for the emergency of the high-$P$ phase. A slightly increasing trend in $T_c$ was observed for both low- and high-$P$ phases with increasing lattice constant $a$ estimated under an ambient pressure.

Fig. 7 (a) presents the X-ray diffraction patterns of $Sr_{0.5}La_{0.5}FBiS_2$ at room temperature under various applied pressures of up to 3.4 GPa. Shifts of the (001) and (004) peaks to higher angles clearly indicate the shrinkage of the lattice along the $c$-axis due to the pressure. In contrast, a relatively smaller shift of the (110) peak was detected, which indicates that the in-plane size remains almost unchanged. Strong peak broadening was observed for the (200) peak above 1.1 GPa, indicating peak splitting due to the lowering of in-plane structural symmetry. Similar peak splitting on the (200) peak was observed for isostructural $LaO_{0.5}F_{0.5}BiS_2$ and $EuFBiS_2$ samples under high pressure[34,35,41], and a resultant structural transition from a tetragonal to monoclinic phase was detected. We noticed that the (200) peak asymmetrically split into two or more peaks, as depicted in Fig. 7 (b). This may be due to the inhomogeneity of applied pressure and the flexible nature of the in-plane structure of $BiS_2$-based compounds. The critical pressure of 1.1 GPa estimated from the XRD corresponds satisfactorily with the $P_c$ estimated from the magnetization measurements.

## Conclusion

Herein, we report on the results of the crystal structure, resistivity, and magnetic susceptibility under pressure for (Sr,$RE$)$FBiS_2$ ($RE$: La, Ce, Pr, Nd, and Sm). The effects of external pressure on magnetization resulted in abrupt increments in $T_c$ up to 10–10.8 K, for samples with $RE$ = La, Ce, Pr, and Nd. Based on analyses of the shielding volume fraction estimated via magnetic susceptibility measurements, we found that two bulk superconductivity phases (low-$P$ and high-$P$ phases) existed for (Sr,$RE$)$FBiS_2$. The common features were that the bulk low-$P$ phase was induced by external pressure, whereas the bulk high-$P$ phase was induced at pressures exceeding a critical pressure. Moreover, at the critical pressure, $T_c$ sharply increased to the high-$P$ phase as our experiment shifted to higher pressures with decreasing $RE$ ionic radius; this implied that both the external as well as chemical pressures were affecting $T_c$. In addition, we investigated the variations in the crystal structure under pressure. By analysing the splitting of the (200) peak, we concluded that the high-$P$ phase was a monoclinic structure. Various bulk superconducting phases with different space group induced via external pressure effects in (Sr,$RE$)$FBiS_2$ need to be further investigated in order to elucidate the mechanisms of superconductivity in Bi$Ch_2$-based systems.

## Methods

Polycrystalline $Sr_{0.5}RE_{0.5}FBiS$ ($RE$: La, Ce, Pr, Nd, and Sm) samples were synthesized by solid state reaction method in an evacuated quartz tube. Powders of $RE_2S_3$ ($RE$: La (99.9%), Ce (99.9%), Pr (99%), Nd (99%), and Sm (99.9%)), $SrF_2$ (99%), Bi (99.999%), and S (99.9999%) were weighed for $Sr_{0.5}RE_{0.5}FBiS_2$. The mixed powder was subsequently pelletized, sintered in an evacuated quartz tube at 700°C for 20 hours, followed by furnace cooling to room temperature. The obtained compounds were thoroughly mixed and ground, then sintered in the same conditions at the first sintering.

The phase purity and the crystal structure of the $Sr_{1-x}RE_xFBiS_2$ ($RE$: La, Ce, Pr, Nd, and Sm) samples were examined by powder synchrotron XRD with an energy of 25 keV ($\lambda$ = 0.49657 Å) at the beamline BL02B2 of SPring-8 under a proposal No. 2019A1101. The synchrotron XRD experiments were performed at room temperature with a sample rotator system, and the diffraction data were collected using a high-resolution one-dimensional semiconductor detector MYTHEN [Multiple mythen system] with a step of $2\theta$ = 0.006°. To investigate the crystal structure of $Sr_{0.5}La_{0.5}FBiS_2$, XRD experiments under high pressure up to 3.4 GPa were performed at room temperature using a Mo-$K_\alpha$ radiation on a Rigaku (MicroMax-007HF) rotating anode generator equipped with a 100 μm collimator. Daphne 7474 was used as a pressure medium.

The crystal structure parameters were refined using the Rietveld method with a RIETAN-FP software[36]. The actual compositions of the obtained samples were analyzed using an energy dispersive X-ray spectroscopy (EDX) on TM-3030 (Hitachi).

The temperature dependence of magnetic susceptibility at ambient pressure and under high pressures were measured using a superconducting quantum interference devise (SQUID) with MPMS-3 (Quantum Design). Hydrostatic pressures were generated by the MPMS high pressure Capsule Cell. The sample was immersed in a pressure transmitting medium (Daphene 7373) covered with a Teflon cell. The pressure at low temperature was calibrated from the superconducting transition temperature of Pb manometer. The electrical resistivity was measured on a GM refrigerator system (Made by Axis) using a conventional four-probe method. For the resistivity measurements, gold wires were connected to the samples with a silver paste.

## Acknowledgments

The authors thank O. Miura for experimental supports. This work was partly supported by JSPS KAKENHI (Grant No. 18KK0076, 15H05886, and 19K15291) and Advanced Research Program under the Human Resources Funds of Tokyo (Grant Number: H31-1) and Nihon University President Grant Initiative.

## Author Contributions

A.Y. prepared all samples and did compositional analysis, resistivity, magnetization measurements. A.Y. measured magnetization under high pressure with the help of R.J. A.M., C.M. and Y.K. conducted powder synchrotron XRD measurements. C.K., K.I. and H.T. conducted the high-pressure XRD measurements. A.Y. wrote the manuscript and prepared all figures. Y.G. and Y.M. provided valuable suggestion for data analysis and corrected the manuscript. All authors contributed to discussion and reviewed the manuscript.

## Additional Information

**Competing Interests**: The authors declare no competing interests.

**Publishing's note** Springer Nature remains neutral with regard to jurisdictional claims in published maps and institutional affiliations.

**Figures**

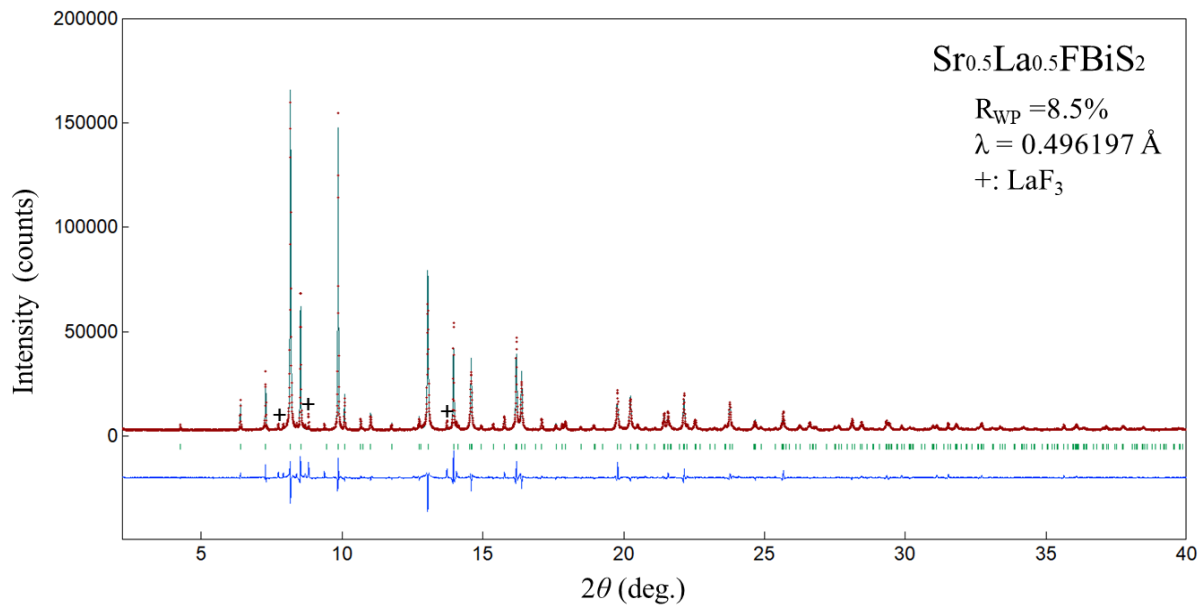

Fig.1 SXRD patterns for $Sr_{0.5}La_{0.5}FBiS_2$. Symbol of + indicates the impurity of $LaF_3$.

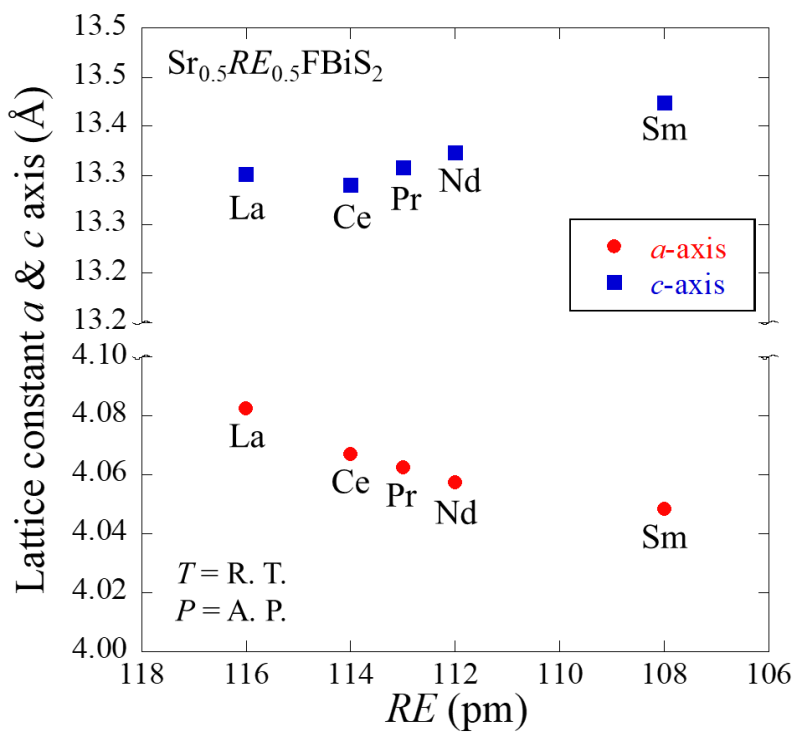

Fig.2 Dependences of the lattice constants $a$ and $c$ as a function of the $RE^{3+}$ ($RE$: La, Ce, Pr, Nd, and Sm) ionic radius.

| Nominal composition | Actual composition |
|---|---|
| $Sr_{0.50}La_{0.50}FBiS_2$ | $Sr_{0.52}La_{0.48}FBi_{1.00}S_{2.00}$ |
| $Sr_{0.50}Ce_{0.50}FBiS_2$ | $Sr_{0.54}Ce_{0.48}FBi_{0.98}S_{2.00}$ |
| $Sr_{0.50}Pr_{0.50}FBiS_2$ | $Sr_{0.56}Pr_{0.41}FBi_{0.99}S_{2.04}$ |
| $Sr_{0.50}Nd_{0.50}FBiS_2$ | $Sr_{0.60}Nd_{0.39}FBi_{0.98}S_{2.03}$ |
| $Sr_{0.50}Sm_{0.50}FBiS_2$ | $Sr_{0.64}Sm_{0.39}FBi_{0.97}S_{2.01}$ |

Table 1 Actual composition (mol%) from the EDX analysis against the nominal composition (mol%). Fluorine amount is regarded as 1.

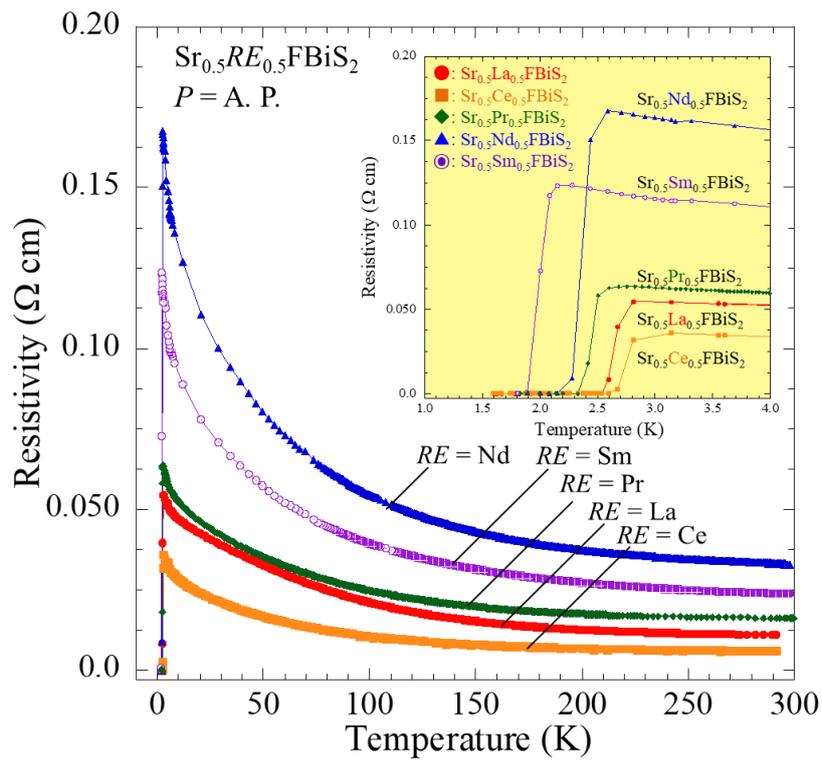

Fig. 3 Temperature dependence of resistivity for $Sr_{1-x}RE_xFBiS_2$ (RE: La, Ce, Pr, Nd, and Sm) at ambient pressure.

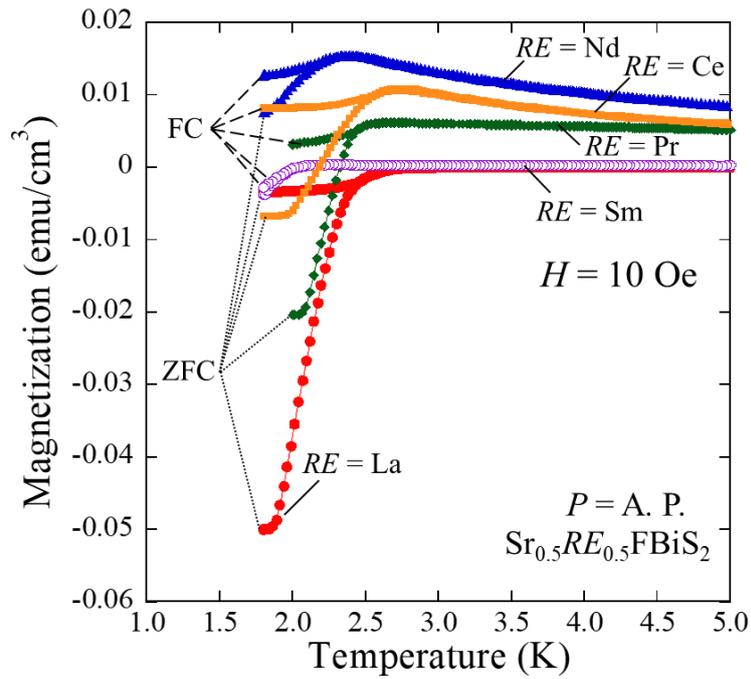

Fig.4 Temperature dependence of magnetization for Sr$_{1-x}$RE$_x$FBiS$_2$ (RE: La, Ce, Pr, Nd, and Sm) at ambient pressure. Dashed and dotted lines indicate a field cooling (FC) and zero field cooling (ZFC), respectively.

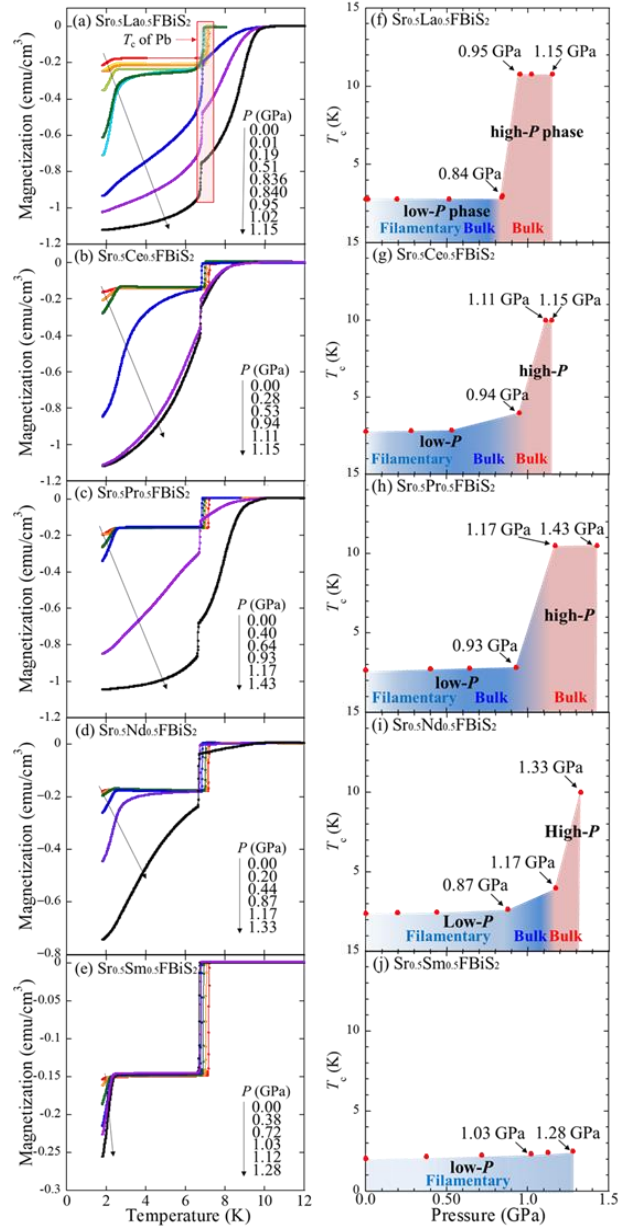

Fig.5 (a-e) Temperature dependence of magnetization under various pressure($P_{La}$ = 0, 0.01, 0.19, 0.51, 0.836, 0.840, 0.95, 1.02, 1.15 GPa; $P_{Ce}$ = 0, 0.28, 0. 53 0.94, 1.11, 1.15 GPa; $P_{Pr}$ = 0.0, 0.40, 0.64, 0.93, 1.17, 1.43 GPa; $P_{Ndr}$ = 0.0, 0.20, 0.44, 0.87, 1.17, 1.33 GPa; $P_{Sm}$ = 0.0, 0.375, 0.716, 1.03, 1.12, 1.28 GPa) for $Sr_{0.5}RE_{0.5}FBiS_2$ (*RE*: La, Ce, Pr, Nd, and Sm), (f-j) Pressure dependence of $T_c$ for $Sr_{0.5}RE_{0.5}FBiS_2$ (*RE*: La, Ce, Pr, Nd, and Sm). Magnetic fields of 10 Oe are applied for all samples. Transition around 7 K indicates the superconducting transition of Pb manometer.

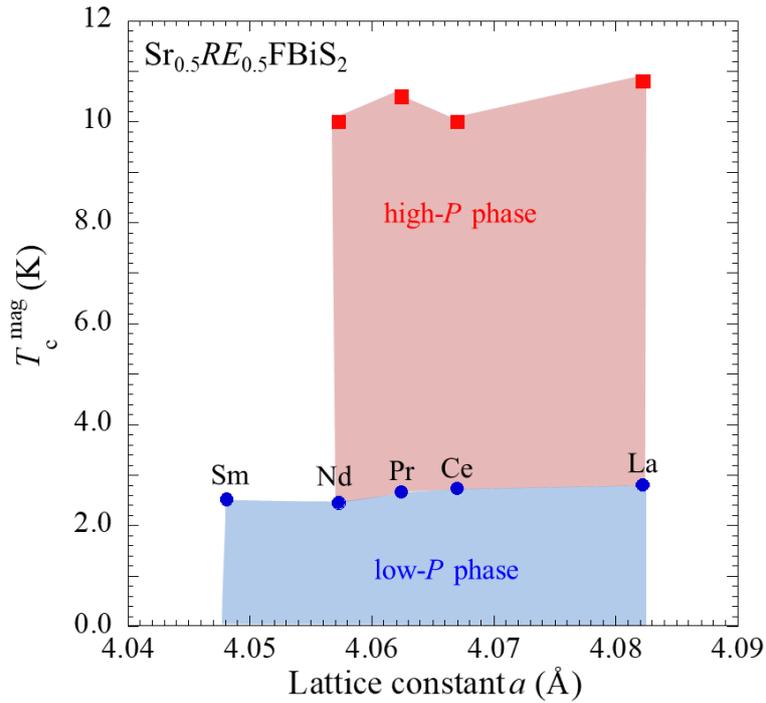

Fig.6 $T_c$ for both low- and high-$P$ phases against the lattice constant $a$ at an ambient pressure.

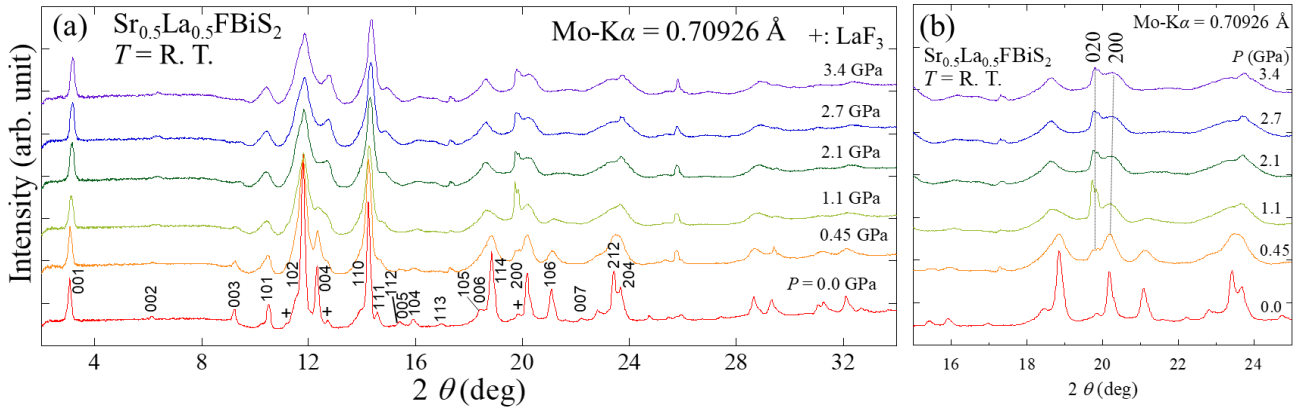

Fig. 7(a) XRD patterns (Mo K$_\alpha$) of Sr$_{0.5}$La$_{0.5}$FBiS$_2$ under various pressure at room temperature. (b) Zoomed XRD patterns near the 020 and 200 peaks.